\documentstyle[version2,prl,aps,twocolumn,epsf]{revtex}
\input psfig.sty

\begin{document}

\widetext
\begin{title}
New Gap Equation for A Marginal Fermi Liquid
\end{title}

\author{Yunkyu Bang}
\begin{instit}
Dept. of Physics, Chonnam National University,  
Kwangju 500-757, Korea
\end{instit}

\vskip .3 in

\begin{abstract}

Assuming a phenomenological self-energy $Im \Sigma(\omega) \sim |\omega|^{\beta}, (\beta=1 $),
which becomes gapped below $T_c$, 
we derived a new gap equation.
The new gap equation contains the effect of the kinetic energy gain upon
developing a superconducting order parameter. 
However, this new kinetic 
energy gain mechanism works only for a repulsive pairing potential leading to a
s-wave state.
In this case, compared to the usual potential energy gain
in the superconducting state as in the BCS gap equation, the kinetic energy gain is 
more  effective to easily achieve a high critical temperature  $T_c$,
since it is naturally Fermi energy scale.
In view of the experimental evidences  of the d-wave pairing state in the hole-doped
copper-oxide high-$T_c$ superconductors, we discuss the implications of our results.

\end{abstract}


\clearpage
\narrowtext

Numerous transport measurements of the high-$Tc$ superconductors\cite{transport} indicate
a very strong quasi-particle  scattering rate ($1/ \tau(\omega)$) up to $ \sim 2000 cm^{-1}$ or even
higher up at the normal state, which is definitely electronic origin and is a source of most of the 
anomalous
normal state properties such as ac/dc conductivity, NMR, Knight shift, ARPES, Raman spectra, etc
\cite{MFL}.
This unique quasi-particle  scattering rate is a finger print of a poor metallic state due to the 
nearness to the Mott insulator.
On the other hand, this strong scattering rate is gapped below 
the superconducting critical temperature $T_{c}$ or 
some higher temperature ($T^{*} >  T_{c}$ ) for most of the 
underdoped compounds,
as indicated by all of the
experiments above mentioned.
Therefore, there is a clear indication that the quasi-particle scattering rate is suppressed as a 
kind of gap develops in the electronic channel and this gap is closely related to the 
superconducting gap\cite{pseudo_gap}.

On the other hand, many theories have been proposed to account for  such  high critical temperature
in the high-$T_c$ cuprates. 
Certainly, it is beyond the scope of this paper to judge the level of success 
of these theories. One of the common ingredients of many theories is that such a strong correlation
effect, as indicated in the many normal state properties, should somehow be involved  in 
the superconducting pairing mechanism. Along this line of thinking, perhaps 
the Interlayer Pair Tunneling
(IPT) theory, proposed early on by P.W. Anderson et. al.\cite{WHA}, 
most actively exploited this idea  to the limit,
such that this theory requires a spin-charge separated 
non-Fermi liquid ground state (called a Luttinger liquid). As a result, the c-axis normal transport is 
dynamically blocked (named as {\it confinement}) but only a momentum conserving 
pair tunneling process is allowed 
via virtual process\cite{chakravarty}. 
More intuitive account of this theory is that the frustrated
kinetic energy along the c-axis in the normal state, 
because of its Luttinger liquid ground state, is liberated in the 
superconducting state and this kinetic energy gain is the main source of the 
high critical temperature \cite{Anderson_book}. 
In this theory, however, there are some of the issues yet 
to be cleared out, such as a realization of the
Luttinger liquid ground state, for example.  
Nevertheless, the idea of the kinetic energy gain as a 
source of superconducting condensation energy seems to be supported with various experiments,
in particular, such as the c-axis ac conductivity and the c-axis penetration depth measurements
\cite{c-axis_exp}.

In this paper, we take a phenomenological approach to the problem how to incorporate the kinetic
energy gain upon entering a superconducting phase in the superconducting pairing mechanism,
not necessarily confining in the c-axis 
kinetic energy. 
We assume that the kinetic energy frustration of the system is represented by a self-energy
$ Im \Sigma(\omega) \sim |\omega|^{\beta}$, and this self-energy becomes gapped upon developing a
superconducting order parameter. We considered mainly $\beta=1$ case since it is the 
most relevant
case for  high-$T_c$ superconductors\cite{transport,MFL}.
We obtained a new gap equation as a natural extension of the BCS gap equation when the condensation 
energy is drained not only from the potential energy but also from the kinetic energy 
as possibly realized with the above assumption.
Oddly enough, the new gap equation shows that the kinetic energy gain works only for a repulsive
pair potential and best for an isotropic  s-wave state. 
For an attractive pair potential it works destructively
and with a strong anisotropy of the order parameter the effect quickly disappears.
We show some numerical results of Tc
as a function of the strength of  kinetic energy frustration 
both for a s-wave and a  d-wave state in a model Tc equation.
The results indicate possible competition between a s-wave  and a d-wave state in 
high-$T_c$ superconductors.
We discuss the implications of our results in view of experiments.
 
We start with a usual BCS Hamiltonian.

\begin{equation}
H = \sum_{k,\sigma} \xi_k c_{k,\sigma}^{\dagger}  c_{k,\sigma} + 
V \sum_{|\xi_{k}|,|\xi_{k^{'}}| < \omega_{D}} 
c_{k,\uparrow}^{\dagger}  c_{-k,\downarrow}^{\dagger} c_{k',\downarrow} c_{-k',\uparrow},
\end{equation}

\noindent
where $c^{\dagger}$ and $c$ are the electron creation and annihilation operators, $\xi_{k}$ is the 
electron energy including chemical potential, $V$ is a two body pair potential.
Now, assuming U(1) symmetry breaking order parameter, 
$\Delta = V \sum_{k^{'};|\xi_{k^{'}}| <  \omega_{D}} < c_{k',\downarrow} c_{-k',\uparrow}>$,
and introducing the Nambu spinor, $\psi^{\dagger} = ( c^{\dagger}_{k,\uparrow}, c_{-k, \downarrow})$,
we can readily obtain the free energy in the Matsubara frequency summation as follows.

\begin{equation}
F(\Delta, \tau) = -\Big[ \frac{\Delta^{*} \Delta}{V} + \tau \sum_{k} \sum_{n} ln 
[\omega_{n}^{2} + (\xi_{k}^{2} + \Delta^{*} \Delta) ]\Big],
\end{equation}
\noindent
where $\tau = k_{B} T$ and $\omega_{n} = \pi \tau (2 n +1)$ is the Matsubara frequencies.

Now, minimizing the above free energy with respect to $\Delta$, i.e., $\delta F / \delta \Delta$ =0,
we obtain the usual BCS gap equation as follows.

\noindent
\begin{equation}
\Delta = - V \tau \sum_{n} \sum_{k, |\xi_{k}| < \omega_{D}} 
\frac{\Delta}{[\omega_{n}^{2} + E_{k}^{2}]},
\end{equation}
\noindent
where $E_{k}^{2} = ( \xi_{k}^{2} + \Delta^{*} \Delta )$.

This is a standard procedure to obtain the self-consistent mean field equation for the order
parameter presumed for a given Hamiltonian. We want now to include the frustrated kinetic energy
and its regaining effect in the superconducting state in the above derivation of the gap equation.
We do this by including the quasi particle renormalization parameter $Z(\omega_{n},\Delta)$
in the above free energy and minimizing it with respect to $\Delta$ {\em with special attention
to the $\Delta$ dependence of  $Z(\omega_{n},\Delta)$.}
We assume the form of a self-energy as follows, which is both experimentally \cite{transport}
and theoretically motivated \cite{rieck}.
\begin{eqnarray}
  Im \Sigma(\omega) &  = &  \alpha |\omega|  ~~\mbox{for}~~ 3 \Delta < |\omega| < \omega_{c} \nonumber \\
               &    &  0           ~~~~~~\mbox{for}~~  |\omega| < 3 \Delta
\end{eqnarray}
\noindent
where $\omega_{c}$ is an ultraviolet cutoff.
For a d-wave gap, more reasonable one would be like
$ Im \Sigma(\omega) \sim \omega^{4} $ for $ |\omega| < 3 \Delta $ .
However, even in that case, we can use the above simplification because it makes no
qualitative change in our main results\cite{w_power}.
The self-energy in the Matsubara frequency, $\Sigma(\omega_{n},\Delta)$,
is obtained by the spectral representation,
\begin{equation}
\Sigma(\omega_{n},\Delta)= -\frac{1}{\pi} 
\Big[  \int^{\omega_{c}}_{-\omega_{c}} - \int^{3 |\Delta|}_{- 3 |\Delta|} \Big] 
\frac{ \alpha |\omega^{'}|}{i \omega_{n} - \omega^{'}} d\omega^{'}
\end{equation}
\noindent
and $\omega_{n}$ is accordingly renormalized.
\begin{equation}
\omega_{n} \rightarrow \tilde{\omega}_{n} = \omega_{n} Z_{n} = \omega_{n} \Big[ 1 + \frac{\alpha}{\pi} 
ln \frac{\omega_{n}^{2} + \omega_{c}^{2}}{\omega_{n}^{2} + (3 |\Delta|)^{2} } \Big].
\end{equation}
With this $\tilde{\omega}_{n}$ in the free energy, Eq.(2), we get
\begin{eqnarray}
- \frac{\delta F}{\delta \Delta^{*}} & = & \frac{\Delta}{V}
+ \tau \sum_k \sum_n \frac{\Delta}{[\omega_n^2 Z_n^2 + E_k^2]} \nonumber \\
& & + \tau \sum_k \sum_n \frac{2 \omega_n^2 Z_n \frac{\partial Z_n}{\partial \Delta^{*}}}
{[\omega_n^2 Z_n^2 + E_k^2]}.
\end{eqnarray}
\noindent
Finally, we obtain the gap equation for a s-wave order parameter, $\Delta =$ const.
\begin{equation}
\Delta = - V \tau \sum_n \sum_{k, |\xi_{k}| < \omega_{D}}
 \frac{[1 - \frac{\alpha}{\pi} \frac{18  \omega_n^2 Z_n}
{[\omega_n^2 +(3 \Delta)^2]}]} {[\omega_n^2 Z_n^2 + E_k^2]} \Delta.
\end{equation}
\noindent
The Tc equation is obtained by taking $\Delta \rightarrow 0$ limit.
\begin{equation}
1= -  V \tau \sum_n \sum_{k, |\xi_{k}| < \omega_{D}} 
\frac{[1 - \frac{\alpha}{\pi} 18 Z_n ^{(0)}]}
{[\omega_n^2 Z_n ^{(0)2} + \xi_k^2]},
\end{equation}
\noindent
where $Z_n ^{(0)} = Z_n (\Delta=0)$.
This is our main result and now let's analyze the above gap equation. 
The term, $\frac{\alpha}{\pi} 18 Z_n ^{(0)}$, in the numerator 
in the right-hand side of Eq. (9) 
is the new term originating from the kinetic energy gain;
otherwise, Eq. (9) is a simple BCS Tc equation with just an ordinary self-energy 
correction, $Z_n$, in the denominator. As well known, if not the new term, the above 
equation can have a solution only for an attractive interaction (i.e. $V < 0$).
However, with the new term,
it can have a solution even with a purely repulsive interaction 
(i.e. $V > 0$) since $\frac{\alpha}{\pi} 18 \sim 2.9$ with $\alpha \sim 0.5$ (which is 
a reasonable value for the high-$T_c$ cuprates\cite{transport}) and $Z_n > 1$ always.
What if  $V < 0$ then ? It is a disaster. The new kinetic energy gain term works
only for a repulsive pair potential but it works destructively for an attractive pair potential.
Therefore I think that the newly found self-energy derived term contains some more physics than the 
kinetic energy gain, as discussed later.

In the case of  d-wave  order parameter, $H_{int}$ and $\Delta(k)$ is defined as follows.
\begin{eqnarray}
H_{int} & = & \sum_{k,k'} V(k,k') c^{\dagger}_{k \downarrow}  c^{\dagger}_{-k \uparrow}
                               c_{k' \uparrow}  c_{-k' \downarrow}, \nonumber \\
\Delta(k) &  = & <\sum_{k'} V(k,k') c_{k' \uparrow}  c_{-k' \downarrow}>.
\end{eqnarray}
\noindent
Here, $V(k,k')$ is a positive definite pair potential 
in momentum space and $\Delta(k)$ is a  sign-changing order parameter of 
d-wave symmetry.
The gap equation can be derived similarly by taking a 
variation  of the free energy, $F_{D-wave}$, with respect to  $\Delta(k)$; it is 
slightly more involved than the s-wave case since the order parameter has a structure.
The result is written as,
\begin{eqnarray}
\Delta(k) & = &
- \tau \sum_n \sum_{k'}  \frac{ V(k,k') \Delta(k') }
                             {[\omega_n^2 Z_n^2 + E_{k'}^2]} \nonumber \\
& & +  \tau \sum_n \sum_{k'} 
\frac{ V(k,k') \frac{\alpha}{\pi} \frac{18 \omega_n^2 Z_n}{[\omega_n^2 +(3 \Delta_{max})^2]}}
{[\omega_n^2 Z_n^2 + E_{k'}^2]} \Delta_{max} \frac{\delta \Delta_{max}}{\delta \Delta(k')}.
\end{eqnarray}
\noindent
Note that the second term contains now $\frac{\delta \Delta_{max}}{\delta \Delta(k')}$, 
which 
has just measure zero since it equals 
$\delta_{k'_{max},k'}~ sgn(\Delta(k'))$; it would be just 1 in the s-wave case.
Therefore, the new  kinetic energy gain term practically has no contribution in 
the d-wave gap equation;
more generally, its effect quickly vanishes for any anisotropic gap $\Delta(k)$
unless $\Delta_{max}$ has a finite support in $k-$ space\cite{d-wave}.
Now in order to gain some more insight about the newly found term, 
we recall the
self-consistency definition of the order parameter,
$\Delta(k) = <\sum_{k'} V(k,k') c_{k' \downarrow}c_{-k' \uparrow}>$.
At first sight, the second term in Eq. (11) appears to be a redundant term.
Although it is not rigorously proved here, 
this extra term doesn't ruin the  self consistency
but actually completes it. We note that  the right-hand side of Eq. (11) is the 
pair susceptibility ( a kind of response function ) with respect to the U(1) symmetry breaking
source, $\Delta(k)$. 
The first term in the right-hand side of Eq. (11) is the usual 
pair susceptibility including the self-energy correction, $Z(\omega_n, \Delta)$.
It is well known that a vertex correction is necessary whenever a self-energy correction 
is included in order to satisfy some continuity equation (or underlying symmetry ) 
for any response function.
Therefore, we can consider the self-energy derived second term as a corresponding vertex 
correction -- it indeed looks like so. And in this sense, the new extra term 
should contain not
only the kinetic energy gain effect but also  some other complicate effect (general backflow).

Now let us show some numerical results of a model Tc equation.
All calculations are performed in two dimensional momentum space
assuming a circular Fermi surface for simplicity of calculations.
In Fig.(1) we show Tc as a function of $\alpha$, 
the strength of  inelastic scattering, both for s-wave and d-wave states
with different values of the coupling constant, $\lambda$, ($\lambda = V N(0)$, 
$N(0)$ is the density of states at Fermi level).
For all calculations, the ultraviolet cutoff scale $\omega_c$ is taken to be  
$0.5eV$ and the BCS potential cutoff $\omega_D$ to be $0.5eV$ 
since the source of the pairing interactions must be an electronic origin.
This choice of parameters is mainly for exemplary purpose, though.
For a s-wave state, the results of  $\lambda = 0.1$ and $0.2$  are shown --
extremely weak repulsive potentials, otherwise
we would get much too high $T_c $.
The message is clear; we can easily achieve several hundreds K of $T_c$ with a very weak
repulsive interaction and it increases with
more kinetic energy frustration (i.e., with  a larger value of  $\alpha$) as expected 
from the Tc equation, Eq. (9). 
For a comparison, we show a d-wave case (solid square).
For a d-wave state, we drop the second term in the gap equation, Eq. (11), as 
explained above and the Tc equation is obtained by taking $\Delta(k) \rightarrow 0$ limit.
For simplicity of numerical calculations, we assume $V(k,k') = V |sin (\phi - \phi')|$ and
$\Delta(k) = \Delta_{max} cos (2 \phi)$ in two dimensional momentum space with a
circular Fermi surface,
and $\phi$ is an angle along the Fermi surface. 
With $\omega_c=0.5eV$ and $\omega_D=0.5eV$ as in the s-wave case, we need $\lambda=1.5$ (an order
of magnitude stronger interaction) in order to get a comparable $Tc$ ($\sim 100 K$)
 for $\alpha \sim 0.4 - 0.6$.
This value of $\alpha$ is the experimental value near optimal doping with $Tc \sim 90 K$ in 
YBCO\cite{transport}. 
$Tc$ quickly decreases with $\alpha$ increasing since now $\alpha$ just
enters the  ordinary pair-breaking self-energy correction in contrast to the s-wave case.
The results show that  if the ground state of the system is a marginal Fermi liquid 
(or even when only some part of Fermi surface is marginal Fermi liquid like)
there is a possible competition between a s-wave state and a d-wave state for the 
experimentally relevant range of $\alpha$, provided 
there is 
a weak repulsive interaction in the s-wave channel\cite{retardation}.

Some remarks and speculations are in order.
First, in this paper we considered only $\beta=1$ case ($Im \Sigma(\omega) \sim |\omega|^{\beta}$). 
we can easily extend our analysis to the other power. 
In general, 
$\delta Z_n (\Delta)/ \delta \Delta \sim \Delta^{\beta}$. 
As a result, if the system is 
a Fermi liquid ($\beta=2$), the kinetic energy gain term  simply drops out of the Tc
equation\cite{w_power}. 
On the other hand, if $\beta$ is any power less than 1, the kinetic energy gain term
becomes singular as $\Delta \rightarrow 0$ and the gap equation becomes
a non-linear equation. In this sense, $\beta=1$ case is  marginal not only for breaking
Fermi liquid\cite{MFL} but also for the superconducting instability.
Second, it is straightforward to extend our formalism to include c-axis dynamics by 
introducing an anisotropy in $\xi_{k}$ and a cylindrical shape of 
order parameter $\Delta(k)$.
The whole formalism goes as the same as in two dimension. There is no kinetic energy 
gain effect for a d-wave order parameter in regard to enhancing Tc 
as the system undergoes from the incoherent state to the coherent state.
The interlayer pair tunneling mechanism by Anderson et. al.\cite{WHA,chakravarty} 
is certainly a different 
approach from ours. 
It is not clear how to reconcile two approaches at present.
Third, as a pure speculation and viewing our numerical results, it may be that the s-wave
instability and the d-wave instability closely compete in the marginal Fermi liquid state
depending
on the potential strength of the s-wave and d-wave channels. 
Then our results  might be  relevant
to the electron-doped high-$T_c$ compound \cite{electron_dope}.

In conclusion, assuming a phenomenological self-energy, 
$Im \Sigma(\omega) \sim |\omega|^{\beta}, (\beta=1 $),
we derived a new gap equation which includes the kinetic energy gain effect.
It is argued that the new gap equation is a natural extension of the BCS gap equation
with a consistent vertex correction 
when the self-energy is a functional of the gap function.
For the corresponding Tc equation, it is also shown that the new kinetic energy gain term 
becomes relevant only when
$\beta \leq 1$,  and $\beta=1$ is the marginal case.
When $\beta=1$, the new Tc equation can provide surprisingly high critical temperature 
for the case of an
isotropic s-wave order parameter with a pure repulsive potential. 
For the d-wave case,
the new kinetic-energy gain term drops from the Tc equation because of the 
anisotropy of the order parameter. Numerical results indicate that there can be a 
competition between the s-wave and d-wave instabilities in a marginal Fermi liquid state depending 
on the strength of pair potential in s-wave and d-wave channels.
In view of the facts that there is a substantial strength of repulsive interaction in s-wave 
channel besides the interaction in d-wave channel in the Hubbard or t-J model and 
its relevance to  high-$T_c$ superconductors,
our results may have some relevance to the electron-doped high-$T_c$ compound.
Considering the unexpected behaviors of the new gap equation, it is desirable to have 
further investigations  of it.

The author thank Jae H. Kim for discussing about related experiments.
He acknowledge the financial support of the Korea Research Foundation, 1996, 
the Korean Science and Engineering Foundation (KOSEF) 
Grant No. 961-0208-048-1,
and the  KOSEF through the SRC program of SNU-CTP.

\psfig{figure=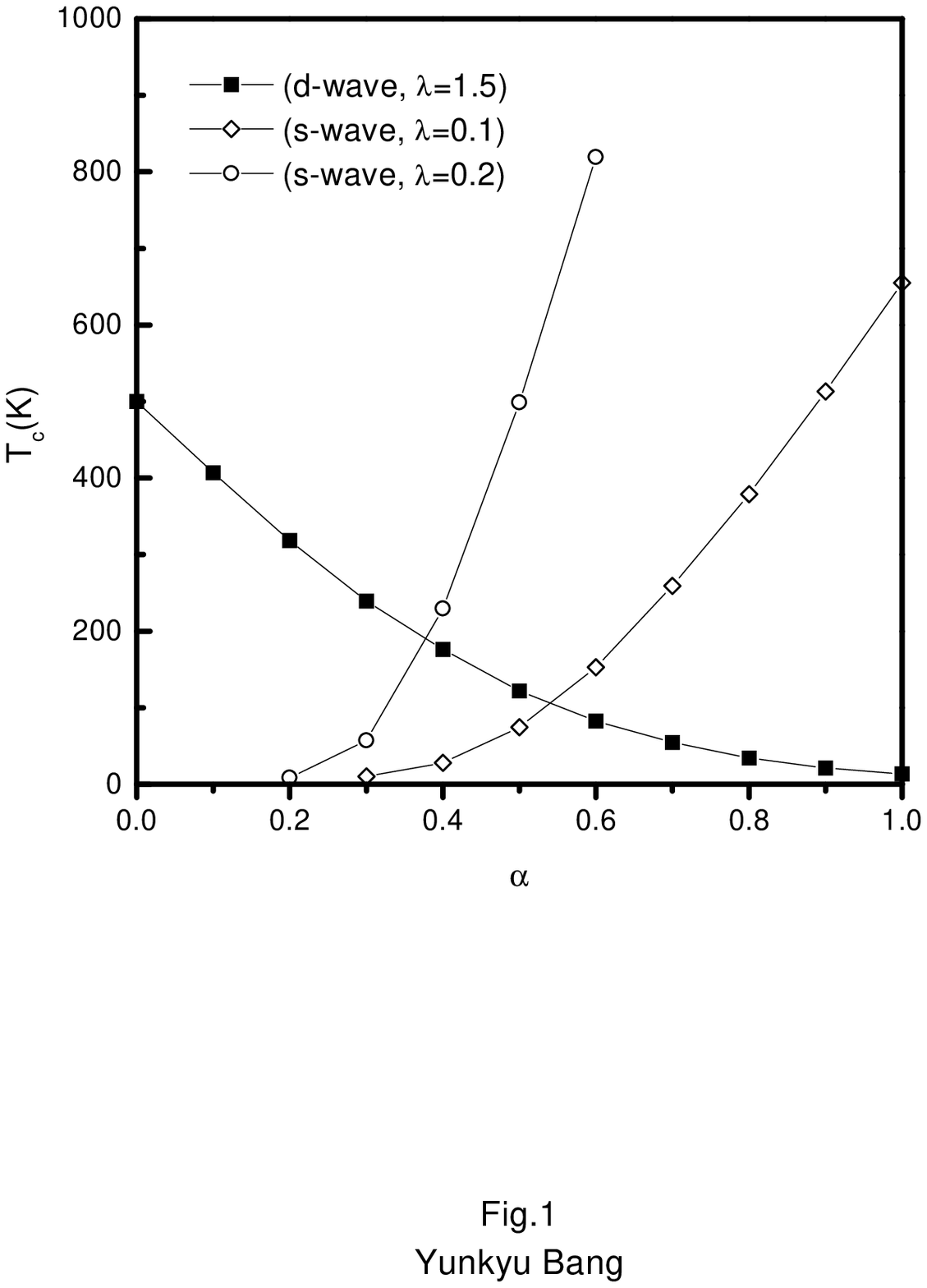,height=16cm}

\figure{Tc vs $\alpha$ ($Im \Sigma(\omega) = \alpha |\omega|$).
For all data, $\omega_D = 0.5 eV$ and $\omega_c = 0.5 eV$; 
solid square (d-wave, $\lambda = 1.5$),
open circle (s-wave,  $\lambda = 0.2$), and open diamond (s-wave,  $\lambda = 0.1$).
\label{fig1}}


\begin{references}
\bibitem[9]{w_power} 
The approximate treatment of $Im \Sigma(\omega)$ 
for $|\omega| < 3 \Delta$ is irrelevant for the Tc equation
as far as  $\gamma > 1$ in  $Im \Sigma(\omega) \sim |\omega|^{\gamma}$ for $|\omega| < 3 \Delta$
since the effect of any higher power term ($\gamma > 1$) drops out in the Tc equation
as taking $\Delta \rightarrow 0$ limit.
However, it makes a difference for the determination of gap value $\Delta(T)$ from the
gap equation.
\bibitem[10]{d-wave}
$\frac{\delta \Delta_{max}}{\delta \Delta(k')}$ in Eq.(11) 
is correct only for the Tc equation because 
of the same reason as explained in [9]. In general, we need a functional derivative, 
$\frac{\delta Z_n (\Delta(k))}{\delta \Delta(k')}$. Then $\Delta(k) < \Delta_{max}$ 
would contribute to the detailed shape of $Im \Sigma(\omega)$ for $|\omega| < 3 \Delta_{max}$ and 
quite generally we can guess  $Im \Sigma(\omega) \sim |\omega|^{\gamma}$ with $\gamma > 1$ for 
any anisotropic order parameter.
Then as argued in [9], only $\Delta_{max}$ part matters for the Tc equation.
A possible exception is the case when $\Delta_{max}$ spans a finite region in momentum
space.
\bibitem[11]{retardation} 
Some might argue that the s-wave channel should be attractive
in real metals if the phonon mediated interaction and the retardation are taken into 
account. The retardation is not compulsory, though. If the system can lower the energy
by developing superconducting order parameter without taking advantage of the retardation,
it will do so.
\end{references}
\end{document}